# Discovery of a 25 parsec-long precessing jet emanating from the old nova GK Persei

Michael M. Shara[1], Kenneth M. Lanzetta[2], James T. Garland[1,3], David Valls-Gabaud[4], Stefan Gromoll[5], Mikita Misiura[6], Frederick M. Walter[2], John K. Webb[7] and Barrett Martin[8]

[1]Department of Astrophysics, American Museum of Natural History, CPW & 79th street, New York, NY 10024, USA

[2]Department of Physics and Astronomy, Stony Brook University, Stony Brook, NY 11794-3800, USA

[3]David A. Dunlap Department of Astronomy and Astrophysics, University of Toronto, 50 St. George Street, Toronto ON MS5 3H4, Canada

[4]Observatoire de Paris, LERMA, CNRS UMR 8112, 61 Avenue de l'Observatoire, 75014 Paris, France

[5]Amazon Web Services, 410 Terry Ave. N, Seattle, WA 98109, USA

[6]Bain & Company, 131 Dartmouth Street, Boston MA 01116, USA

[7]Institute of Astronomy, University of Cambridge, Madingley Road, Cambridge CB3 0HA, United Kingdom

[8]Antioch University, 2400 3rd Avenue, Seattle WA 98121

**Classical nova eruptions result from thermonuclear-powered runaways[1,2] in, and ejection of, the hydrogen-rich envelopes of white dwarf stars accreted from their close binary companions[3,4]. Novae brighten to $10^4$ - $10^6$ solar luminosities, and recur thousands of times over their lifetimes[5,6] spanning several billion years. Between eruptions, mass transfer from the donor star to the white dwarf proceeds via an accretion disk[7] unless the white dwarf possesses a strong magnetic field[8,9] which can partially or totally disrupt the disk[10]. In that case, accretion is focused by the white dwarf's magnetic field towards its magnetic poles. Optical spectroscopy[11,12,13,14] and interferometric radio maps[15] demonstrate the presence of bipolar jets, typically arcseconds in angular size, and orders of magnitude smaller than one parsec in linear size, during the days to months after nova eruptions. These jets expel collimated matter from the white dwarfs in nova binary stars, but well-resolved images of them are lacking. Here we report the Condor telescope's detection of a hitherto unknown, highly resolved braided jet, three degrees (at least 25 parsecs) in length. The jet originates at the white dwarf of the old nova GK Persei (nova Per 1901 CE). It precesses on a ~ 3600 yr timescale, and must be at least 7200 years old. Detected across four decades of wavelength, the jet's ultimate energy source is likely the strong accretion shocks near the white dwarf's magnetic poles.**

**Subject terms: Stars  Novae  Jets**

The highly fragmented ejecta of GK Per (nova Persei 1901)[16,17,18,19] are the largest known for any 20th century nova, spanning 0.2 pc (1.5 arcmin in angular size). Condor Array Telescope[20] g' and r'-band images of a 5 x 2.5 deg field encompassing GK Per reveals a hitherto unknown, 100-fold larger structure, shown in Figure 1: a 3-degree long corkscrew-shaped ``jet''. Closeups of several segments of the corkscrew structure, shown in Figure 2, show it split into two components, demonstrating that it is braided.

A search for this feature at other wavelengths yields positive detections across four decades of the electromagnetic spectrum, from 0.23 μm to 2096 μm. The mosaic of Figure 3 shows that the corkscrew is visible in Planck, Akari, IRIS, WISE, Palomar DSS, Condor and GALEX near-UV images. It is noticeably absent only in the 2MASS near-infrared and GALEX Far-UV passbands. The corkscrew is most obvious in the visible-NUV and the far infrared to microwave parts of the spectrum. As in protostellar jets[21], thermal and non-thermal emission are both likely to be emission mechanisms[22].

A star-subtracted image of the jet is shown in Figure 4 (see Methods), which provides evidence for condensations near the jet's edges that are reminiscent of shock structures seen in protostar ejecta[23].

GK Per is a well characterized[24] binary star with white dwarf and subgiant masses of $M_{WD}$ = $1.03^{+0.16}_{-0.11}$ Msun and $M_{SG}$ = $0.39^{+0.07}_{-0.06}$ Msun; a white dwarf radius[25] $R_{WD}$ = 0.0075 Rsun; a white dwarf spin period[26] $P_{spin}$ = 351.33 s; an orbital period $P_{orb}$ = 1.996872 ± 0.000009 d[24], and a consequent orbital semimajor axis a = 3.4 x $10^6$ km[7]. The contribution of the centrifugal acceleration to the local effective gravity at the equator of the white dwarf yields its oblateness[27] B = [ 2 *π *Rwd / $P_{spin}$]$^2$ / $GM_{WD}/R_{WD}$ = 3 x 10$^{-4}$. If the white dwarf is an oblique rotator (see Methods) then its precession rate[28] under the tidal influence of the subgiant companion is dψ/dt ≅ 1.5 [$GM_{SG}$ / a$^3$ ] [B $P_{orb}$ / 2π ] = 0.1 deg/yr, so its precession period is ~ 3600 yr.

If the terminal ejection velocity of jet-ejected material is ~ half[29] the WD escape velocity (i.e. ~3500 km/s), then a full 360° turn of the corkscrew would correspond to a physical length of ~ 3500 km/s x 3600 yr = 12.6 pc. The projected angular size of the 360° loop of the corkscrew closest to GK Per is ~ 45' which, at the 434 pc distance of GK Per[30] corresponds to 5.7 pc. The orbital inclination[24] of GK Per is 67° ± 5°, so the de-projected length of a single 360° loop is ~14.5 pc, in excellent agreement with the above prediction. The observed length of the corkscrew is at least twice that size, so the corkscrew feature's age must be greater than twice the precession period, i.e. > 7200 yr.

The Condor and GALEX images yield UV + visible surface brightnesses of the corkscrew of ~ 4 x 10^-15 erg/s/cm^2/arcsec^2. The full corkscrew covers ~ 0.4 degree^2 of sky, implying a jet luminosity of order 10 Lsun. GK Per's distance and dwarf nova outburst apparent magnitude $m_V$ ~ 10 correspond to an intermittent white dwarf accretion rate and binary system luminosity of ~3 x 10^-9 Msun/yr and 15 Lsun, respectively, barely sufficient to drive mass flows that match the jet's power. However the mass transfer rates from dwarf[6] and giant[31] donors in cataclysmic binaries vary by orders of magnitude on timescales of millenia, and the subgiant donor's rate in GK Per likely does the same. If that rate is of order 10^-8 Msun/yr for much of the time between GK Per nova outbursts, then the persistence of the GK Per jet over millenia can be powered by accretion energy and mass outflows released near the precessing white dwarf's poles.

**Corresponding Authors**

MS is the corresponding author. Correspondence and requests for materials should be addressed to [mshara@amnh.org](mailto:mshara@amnh.org) .

**Acknowledgements**

We are deeply grateful to Michael Hensley and Diana Casas-Hensley for their strong support in maintaining the Condor Array Telescope. MMS and JTG acknowledge the support of NSF award 2108234. KML is supported by the National Science Foundation under grants 1910001, 2107954, and 2108234. This work has made use of data from the European Space Agency (ESA) mission *Gaia* (https://www.cosmos.esa.int/gaia), processed by the *Gaia* Data Processing and Analysis Consortium (DPAC) (https://www.cosmos.esa.int/web/gaia/dpac/consortium). Funding for the DPAC has been provided by national institutions, in particular the institutions participating in the *Gaia* Multilateral Agreement. It also made use of the software program StarNet v2 produced by co-author Mikita Misiura. MS thanks Tomek Kamiński for a valuable discussion about GK Per's molecular gas.

**Author Contributions**

All authors shared in formulating the ideas underlying the writing of this paper, and each participated in writing sections of it. MS chose the target, specified the filters used in the study and wrote the initial text; KL carried out the imaging observations; KL and SG produced the pipeline that reduced the images; MM produced the StarNet software; and KL, SG, JG, and DVG produced the figures.

## Author Information

Reprints and permission information is available at www.nature.com/reprints.

## Competing Interests

The authors declare that they have no competing financial interests.

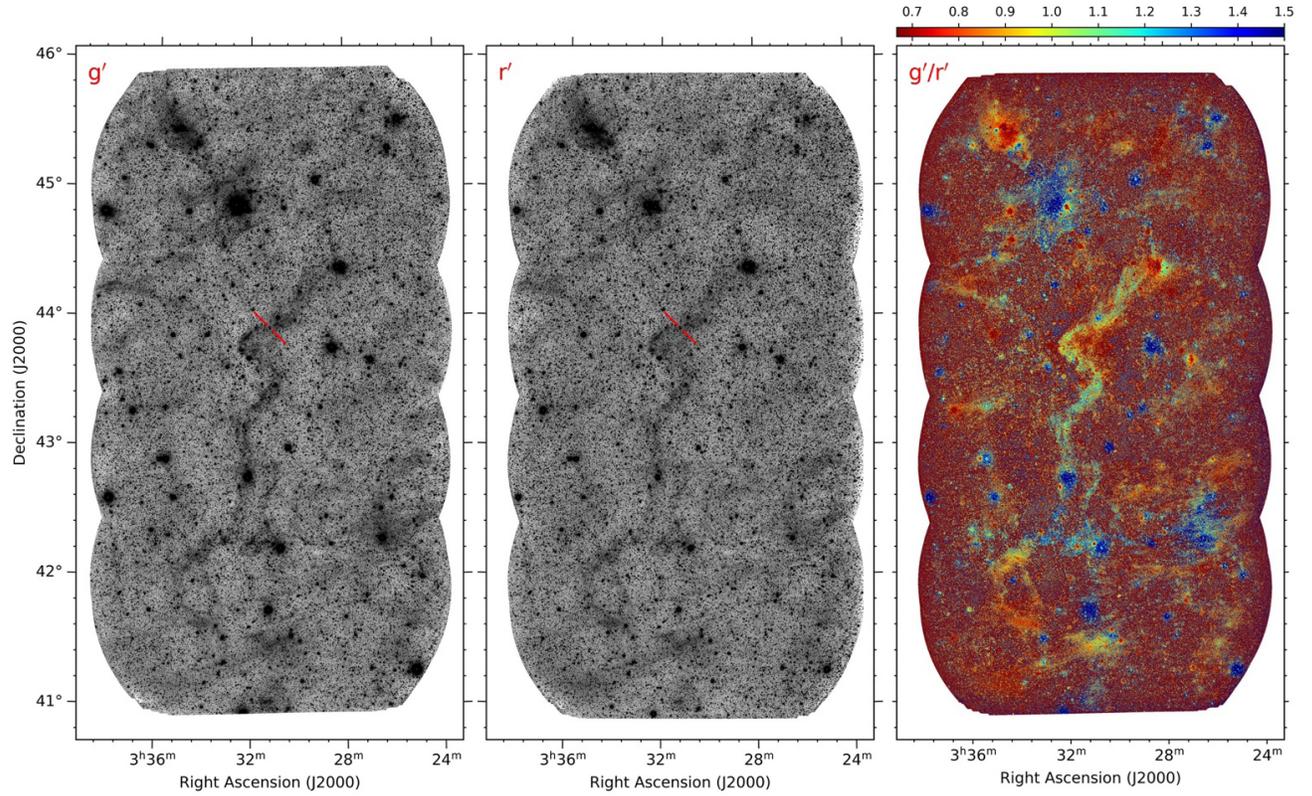

Figure 1: Sloan g', r' and the flux ratio g'/r' images of the precessing jet emerging from the old nova GK Per. The nova is indicated with red lines in the left and central images. The bar above the rightmost (g'/r') image is the scale of the color of the jet along its length.

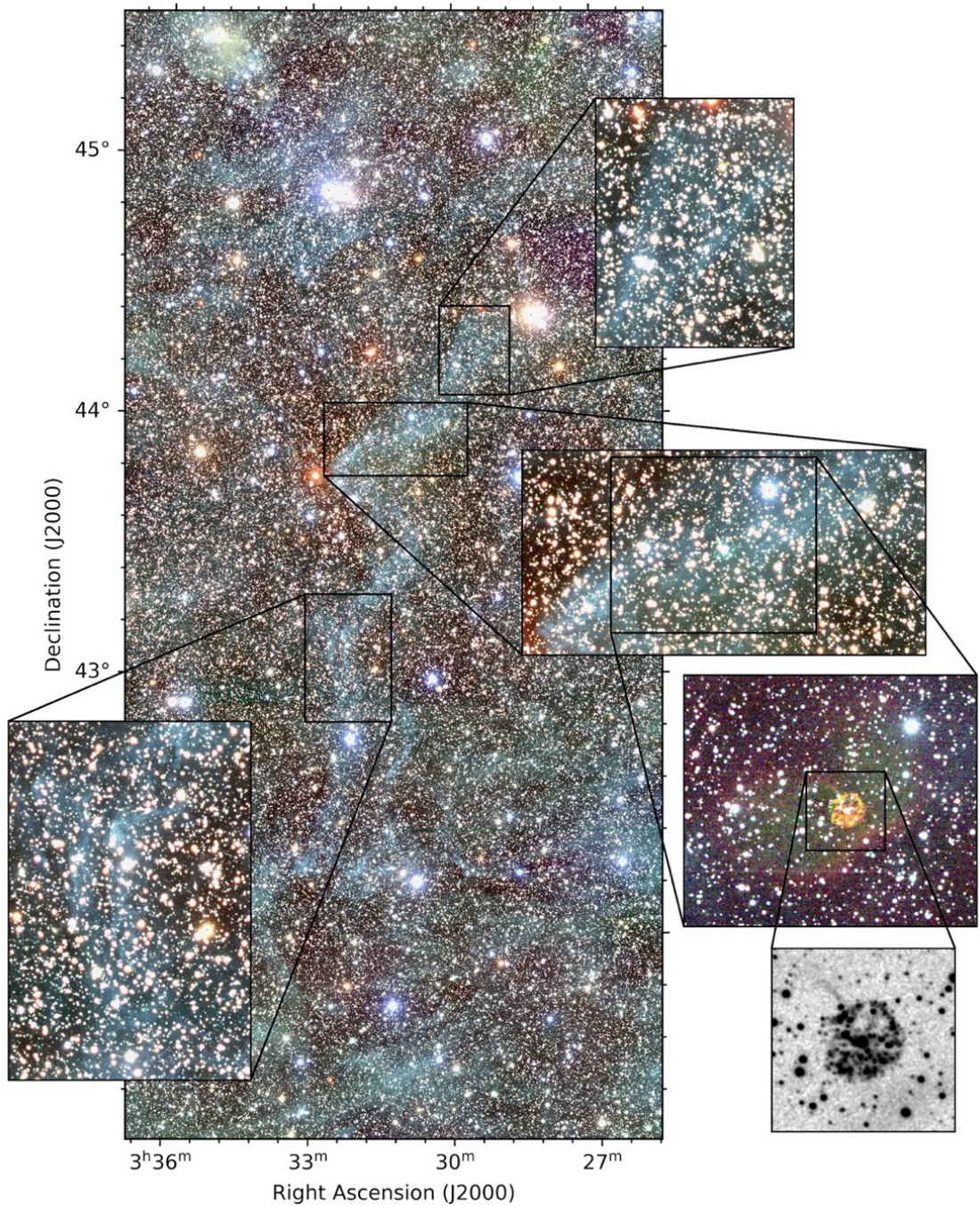

**Figure 2:** Area surrounding GK Per in Condor i', r', and g' images in red, green, and blue, respectively. The images have been smoothed via a Gaussian kernel (σ = 2.5 pixels = 2.1") and are displayed with asinh scaling. Inset close-ups highlight the northern and southern bifurcations as well as the central nebulosity around GK Per. A zoomed-in color image with Condor H α + [N II], [O III], and He II as red, green, and blue shows the narrowband nebulosity nearby GK Per, including the ridge to the SW of the star; and a further zoomed-in inset of H α + [N II] reveals the 1901 nova ejecta in detail.

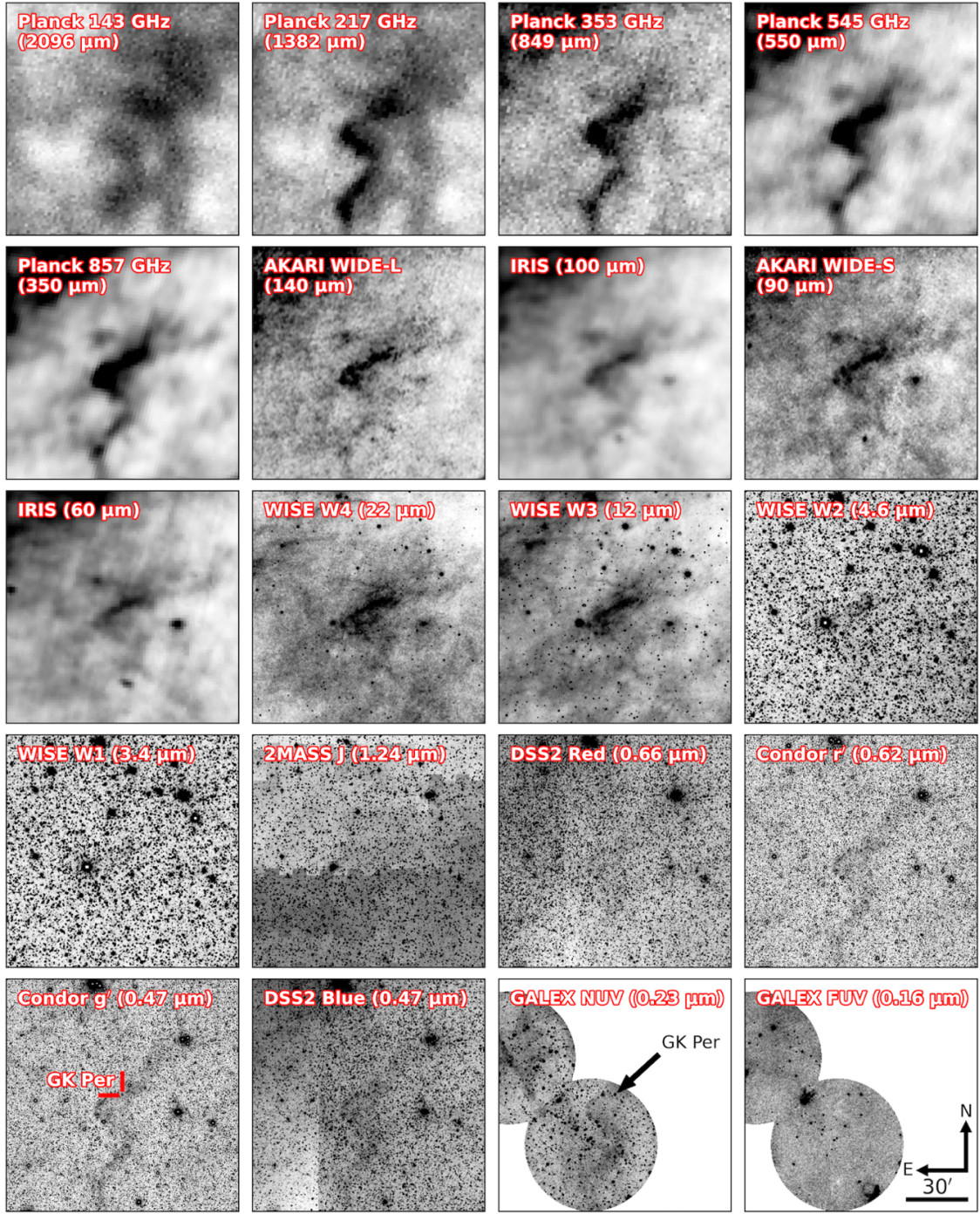

Figure 3: 2x2 degree images of the precessing jet of GK Per from the Planck, AKARI, IRIS, Wise, 2MASS, Palomar and GALEX sky surveys, covering over four decades in wavelength. The central wavelength of each image, and the satellite which took it are indicated at the top of each image.

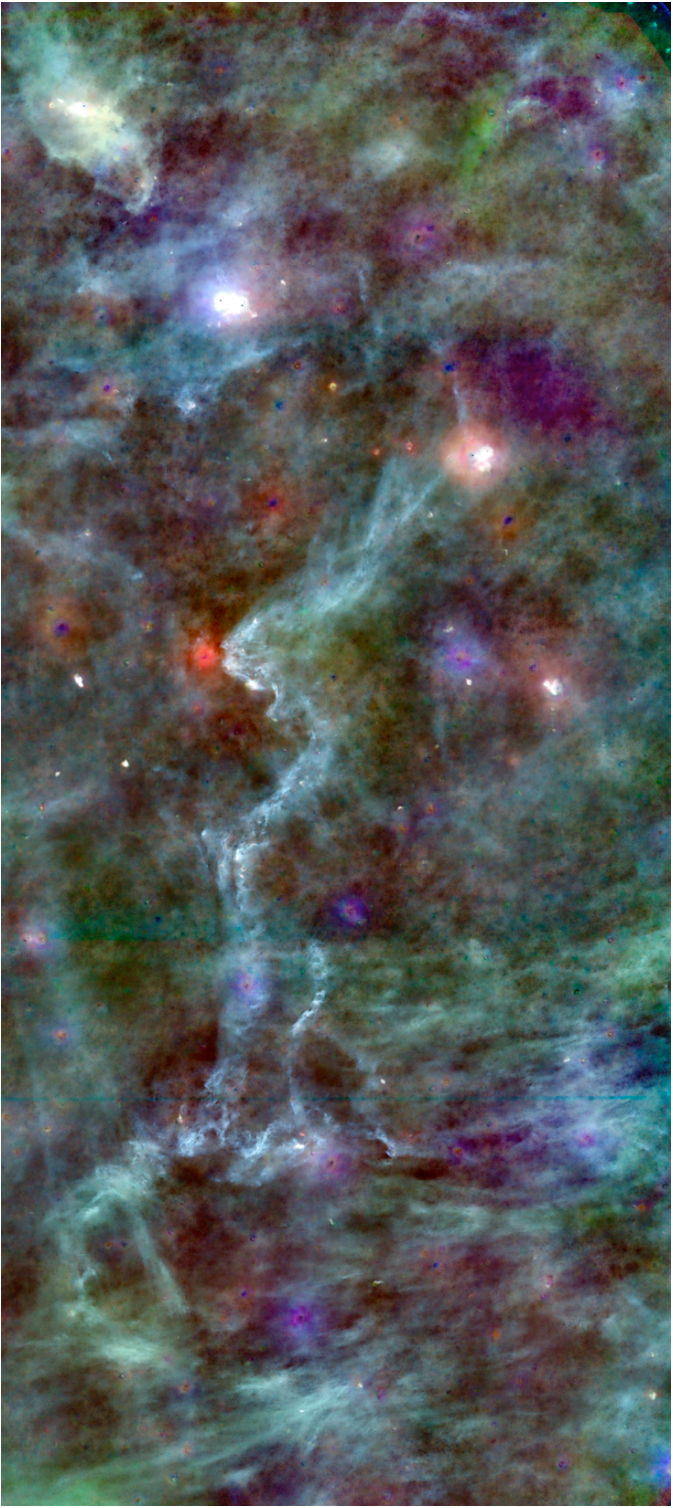

**Figure 4. The ``star-subtracted" GK Per precessing jet. The g', r' and i' images are assigned blue, green and red colors. The StarNet V2 algorithm noted in the Methods section has been used to digitally remove almost all the stars in the field of view.**

**Data Availability**

All data pertaining to these observations is available upon reasonable request from MS.

## Methods

**The Telescope**

All the images presented here were obtained with the Condor Array Telescope[20], which is comprised of six fast (f/5) apochromatic refractors with 180 mm objectives. Each refractor is equipped with a 9576 x 6388-pixel CMOS camera covering 2.3 x 1.5 deg$^2$, and Sloan g', r' and i' filters. The telescope is located at a very dark site in southern New Mexico, USA. Astrometric calibration, field flattening, background subtraction, and photometric calibration have all been described[20].

**The Data**

GK Per and its environs were observed with the Condor Array Telescope through Sloan g', r' and i' filters on multiple nights between 15 November 2022 and 02 February 2023. Four partially overlapping pointings covered a region 2.3 x 5 degrees in extent. The telescope was dithered by a random offset of ~ 15 arcmin between each exposure. Because different filters were used on different telescopes at different times, we define the "reach" of an observation obtained by Condor observations as the product of the total objective area and the total exposure time devoted to the observation. Since Condor consists of six individual telescopes, each of objective area 0.0254 m$^2$, a one-second exposure with one telescope of the array yields a reach of 0.0254 m$^2$ s, and a one-second exposure with the entire array (i.e. with all six telescopes) yields a reach of 6 x 0.0254 m² s = 0.153 m² s. The reaches of the g', r' and i' images centered on GK Per were 9603, 11,054 and 4137 m² s, while those of the other three fields, north and south of GK Per, averaged 5000, 5000 and 1500 m² s.

**The scales of stellar jets**

Jets are ubiquitous in astrophysics, appearing at the centers of accretion disks where infalling matter is accelerated and collimated hydromagnetically[32,33,34]. A comparison of the GK Per jet with two of the best-known jets launched from compact objects provides useful perspective.

R Aqr is a symbiotic binary system consisting of an Asymptotic Giant Branch mira variable, a hot companion (probably a white dwarf) with a spectacular jet outflow, and an extended emission line nebula[35,36]. At a Gaia-determined distance of 386 +/- 50 pc[37], the 2.5' arcmin long jet of R Aqr extends 0.28 pc from end-to-end. This is ~100X smaller that the jet of GK Per.

SS 433 is a high mass X-ray binary whose compact object (a neutron star or a black hole) generates a pair of well-collimated relativistic jets[38,39] which precess on a cone of half-angle $\theta \simeq 20°$ with a period P = 163 d. The radio jets subtend ~5" which, at the 5.5 +/- 0.2 kpc distance of SS 433 correspond to 0.13 pc. This is ~ 200X smaller than the GK Per jet.

**The ``star-removed" Jet**

We have used the software package *StarNet version 2*[41] (a convolutional residual net with encoder-decoder architecture) to remove stars from the g', r' and i' images described above, and then co-added these three images to produce Figure 4. Notable features in

Figure 4 include shock-like structures near edges of the jet, like those seen in the jets of Protostars[23,42].

**Oblique Rotator in GK Per and Cataclysmic Binaries**

For the GK Per jet to precess, the obliquity of its white dwarf's axis of rotation must be nonzero. While the jet doesn't have sharp edges, its maximum angular extension (measured as the angle from GK Per to the first and second maximum extensions of the corkscrew) is close to 70 degrees. The obliquity of the GK Per white dwarf is thus ~ 35°. Other well-known cataclysmic binaries that are oblique rotators include DQ Her-type stars[43] and AE Aqr[44].

**Gas Surrounding GK Per**

Extended emission seen in IRAS 100 μm images[45], as well as HI and narrowband optical imaging[46] led to the suggestion that the explosion of the nova into a planetary nebula could account for the extent and morphology of the observed ejecta. This model was questioned[47] on the basis of a 115 GHz $^{12}$CO survey which suggested that at least some of the elongated dust emission detected by IRAS is in fact contained in normal interstellar cirrus clouds.

**Bifurcated CO gas stream near GK Per**

Cool molecular CO gas, closely following the spatial distribution of the jet, has been observed from ~500" distant to within ~ 100" of GK Per, both Eastward and Westward of the binary[49]. A highly significant gap, centered precisely on GK Per, is seen in the 115 GHz IRAM 30-meter telescope images of the stream of CO-emitting gas (see Figure 5 of reference 49). This gap is also visible, albeit at lower resolution, in the Planck 217 GHz image of Figure 2 of this paper. The centering of the gap exactly on GK Per is much too precise to be a coincidence. Rather, it demonstrates that the molecular gas (and the jet) are at the same distance as GK Per, and that the harsh radiation environment of GK Per is responsible for photo-dissociating CO molecules within ~ 0.5 pc of the accreting binary.

**Methods References**